\documentclass[twocolumn,prc,showpacs,preprintnumbers,unsortedaddress,amsmath,amssymb,floatfix]{revtex4}
\usepackage{graphicx,color}
\usepackage{bm}
\usepackage{graphicx}
\usepackage{CJK}

\usepackage{comment}
\usepackage{color}
\usepackage{slashed}
\usepackage{ulem}

\newcommand{\delete}{\bgroup\markoverwith{\textcolor{red}{\rule[0.5ex]{2pt}{1pt}}}\ULon}

\newcommand{\be}{\begin{equation}}
\newcommand{\ee}{\end{equation}}
\newcommand{\bea}{\begin{eqnarray}}
\newcommand{\eea}{\end{eqnarray}}

\begin{document}
\begin{CJK*}{GBK}{song}

\title{Spin-isospin Response in Finite Nuclei from an Extended Skyrme Interaction}
\author{Peiwei Wen$^{1,2}$, Li-Gang Cao$^{1,3,4,5}$, J. Margueron$^{6}$, H. Sagawa$^{7,8}$}

\address{${}^1$Institute of Modern Physics, Chinese Academy of
Sciences, Lanzhou 730000, China}

\address{${}^2$ University of Chinese Academy of
Sciences, Beijing 000049, China}

\address{${}^3$ State Key Laboratory of Theoretical Physics, Institute of Theoretical Physics, Chinese Academy of Sciences, Beijing 100190, China}

\address{${}^4$ Kavli Institute for Theoretical Physics China, CAS, Beijing 100190, China}

\affiliation{${}^5$ Center of Theoretical Nuclear Physics, National
Laboratory of Heavy Ion Accelerator of Lanzhou, Lanzhou 730000, P.R.
China}

\address{${}^6$
Universit\'e de Lyon, Universit\'e Lyon 1 and CNRS/IN2P3,
Institut de Physique Nucl\'eaire de Lyon,
4 rue Enrico Fermi, 69622 Villeurbanne, France}

\affiliation{${}^7$Center for Mathematics and Physics, University of
Aizu, Aizu-Wakamatsu, Fukushima 965-8580, Japan}

\affiliation{${}^8$RIKEN, Nishina Center, Wako, 351-0198 ,Japan}

\begin{abstract}
The magnetic dipole (M1) and the Gamow-Teller (GT) excitations of
finite nuclei have been studied in a fully self-consistent
Hartree-Fock (HF) plus random phase approximation (RPA) approach by
using a Skyrme energy density functional with spin and spin-isospin
densities. To this end, we adopt the extended SLy5st interaction
which includes spin-density dependent terms and stabilize nuclear
matter with respect to spin instabilities. The effect of the
spin-density dependent terms is examined in  both the mean field and
the spin-flip excited state calculations. The numerical results show
that those terms give appreciable  repulsive contributions to the M1
and GT response functions of finite nuclei.
\end{abstract}


\maketitle

\section{Introduction \label{intro}}

The properties of spin asymmetric matter are still very difficult to
access experimentally since the ground states of nuclei have a weak
or an almost zero spin-polarization: even-even spherical nuclei are
not spin-polarized, while their closest odd nuclei can be weakly
spin-polarized by the last unpaired nucleon, but with a moderate
impact on the ground-state energy~\cite{marg2}. In well deformed
nuclei, ground-state spin and parity assignments are still difficult
to predict globally~\cite{Bonneau2007}. It is therefore difficult to
probe the nuclear interaction in spin and spin-isospin channels from
the ground-state properties of nuclei. However, in the excitation
spectra of nuclei, some collective modes can provide a unique
opportunity to explore the nuclear interactions in spin and
spin-isospin channels~\cite{Harakeh,Ost92,Fuj11}.  The M1 and GT
excitations are the most common collective modes of spin and
spin-isospin types in nuclei. These modes have been extensively
studied during the last decade and much information on spin and
spin-isospin excitations becomes now available~\cite{Fuj11,Key10}.
They are of interest not only in nuclear physics but also in
astrophysics. They play, for instance, an important role in
predicting $\beta$ decay half-lives of neutron rich nuclei involved
in the r-process of the nucleosynthesis~\cite{Bor06}. In
core-collapse supernova, the GT transitions of pf-shell nuclei give
an important contribution to the weak interaction decay rates that
play an essential role in the core-collapse dynamics of massive
stars~\cite{Fre06,Ich06,Fan12}. The neutrino-induced nucleosynthesis
may take place via GT processes in neutron-rich
environment~\cite{She99}. For neutrino physics and double $\beta$
decay, accurate GT matrix elements are necessary to understand the
nature of neutrinos~\cite{Tad87}.

In  the beginning of the 1980s, GT experiments made great progress
 when the (p,n) facility at the Indiana University Cyclotron
Facility became operational. In 1981, the Skyrme SGII interaction
was design to give, for the first time, a detailed description of
the GT data~\cite{Gia81}. Some other Skyrme interactions, such as
SLy230a \& SLy230b~\cite{Cha97}, SLy4 \& 5~\cite{Cha98},
SkO~\cite{Rei99}, and more recently SAMi~\cite{Roca12}, have been
determined with a special care of the spin and spin-isospin
properties of nuclear matter and  nuclei. Calculations of GT within
the relativistic framework were done  more
recently~\cite{Paa04,Liang08}. The relation between the spin or the
spin-isospin excitations and the central part of the nuclear
interaction is however not a one-to-one relation and other effects
should be considered such as the spin-orbit splitting of the
single-particle states and the residual spin-orbit interaction in
the RPA calculations~\cite{Ben02}.

Recently an extension of the Skyrme interaction, including
spin-density and spin-isospin density dependent terms, was proposed
by some of the present authors~\cite{marg1,marg2}. At variance with
predictions in nuclear matter of $ab$ $initio$ methods based on
realistic bare interactions~\cite{Fan01,Via02,Via021,Bom06}, most of
the standard Skyrme interactions predict spin or spin-isospin
instabilities beyond the saturation density of nuclear
matter~\cite{Margueron2002}. The additional parameters of the
extended Skyrme interaction were therefore adjusted to reproduce the
results given by microscopic G-matrix calculations better. The
extension of the Skyrme interaction was designed  to keep the
simplicity of the standard Skyrme interaction and to remove the
ferromagnetic instability or to shift it to larger density.

The extended spin-density dependent terms can improve the properties
of the Skyrme energy density functional in spin and spin-isospin
channels by adding the weak repulsive effect. For example, the
dimensionless Landau parameter $G_0^\prime$ is increased by about
0.3 for three interactions SLy5~\cite{Cha98}, LNS~\cite{Cao06} and
BSk16\cite{Cha08}. In Refs.~\cite{marg1,marg2}, the authors explored
the effect of spin-density dependent terms on the response functions
and the mean free path of neutrinos in nuclear matter as well as the
ground state properties of finite odd nuclei. The model proposed in
Ref.~\cite{marg1,marg2} was  constrained by microscopic G-matrix
predictions in uniform matter. It will be  quite interesting to
investigate the effect of the proposed extension of the Skyrme
interaction  for  the spin and spin-isospin excitations of finite
nuclei. In the present work, we study the contribution of
spin-density dependent terms to the M1 and GT excitations in finite
nuclei $^{90}$Zr and $^{208}$Pb with a fully self-consistent HF plus
RPA framework~\cite{colo13}. The SLy5 Skyrme parameter set is
employed in our calculations by adding  the spin-density dependent
terms. The new parametrization is called as SLy5st, which is the
same  used in Refs.~\cite{marg1,marg2}. In present study we switch
on and off the spin-density dependent terms in ground states and
excited states calculations to see how much they affect the spin and
spin-isospin response functions in finite nuclei.

This paper is organized as follows. In Sec. II we will briefly
report the theoretical framework of the RPA based on the Skyrme
interaction and its extension. The results and discussion are
presented in Sec. III. Section IV is devoted to the summary and
perspective for future.

\section{Formula}

We adopt the standard form of Skyrme interaction with the notations
of Ref.~\cite{Cha98}. The two nucleons are
interacting through a zero-range, velocity-dependent and density-dependent Skyrme
interaction  with space,
 spin and isospin variables $\bm{r}_i$, $\bm{\sigma}_i$ and $\bm{\tau}_i$  which reads 
\cite{Cha98}:
\begin{eqnarray}
V({\bm r}_1, {\bm r}_2) &=& t_0(1+x_0 P_{\sigma}) \delta({\bm r})
\nonumber \\ &+& \frac{1}{2} t_1 (1+x_1P_{\sigma})[{\bm
P}'^{2}\delta({\bm r})+
\delta({\bm r}){\bm P}^{2}] \nonumber \\
&+& t_2(1+x_2P_{\sigma}){\bm P}'\cdot\delta({\bm r}){\bm P}
\nonumber \\&+&
\frac{1}{6} t_3 (1+x_3P_{\sigma})\rho^{\alpha}({\bm R})\delta({\bm r})\nonumber\\
&+& iW_0 ({\sigma}_1+{\sigma}_2)\cdot[{\bm P}'\times\delta({\bf
r}){\bm P}]~, \label{eq2.1-1}
\end{eqnarray}
where $\bm{r}=\bm{r}_1-\bm{r}_2$,
$\bm{R}=\frac{1}{2}(\bm{r}_1+\bm{r}_2)$,
$\bm{P}=\frac{1}{2i}(\bm{\nabla}_1-\bm{\nabla}_2)$, $\bm{P}'$ is the
hermitian conjugate of $\bm{P}$ (acting on the left),
$P_{\sigma}=\frac{1}{2}(1+\bm{\sigma}_1\cdot\bm{\sigma}_2)$ is the
spin-exchange operator, and $\rho=\rho_n+\rho_p$ is the total
nucleon density. Within the standard formalism, the total binding energy
of a nucleus can be expressed as the integral of a Skyrme density
functional~\cite{Cha98}, which includes the kinetic-energy term
$\mathcal{K}$, a zero-range term $\mathcal{H}_0$, the
density-dependent term $\mathcal{H}_3$, an effective-mass term
$\mathcal{H}_{eff}$, a finite-range momentum dependent term $\mathcal{H}_{fin}$, a
spin-orbit term $\mathcal{H}_{so}$, a spin-gradient term
$\mathcal{H}_{sg}$, and a Coulomb term $\mathcal{H}_{Coul}$.

The Skyrme interaction has been extended to include spin-density
dependent terms which can improve the properties of the energy
density functional in the spin and spin-isospin
channels~\cite{marg1}. That is \bea
V^\mathrm{add.}(\bm{r}_1,\bm{r}_2)&=&
\frac{1}{6}t_3^s(1+x_3^sP_\sigma)[\rho_s(\bm{R})]^{\gamma_s}\delta(\bm{r})
\nonumber\\
&+&\frac{1}{6}t_3^{st}(1+x_3^{st}P_\sigma)[\rho_{st}(\bm{R})]^{\gamma_{st}}\delta(\bm{r}),
\label{eq:addint} \eea
 where  $\rho_s=\rho_\uparrow-\rho_\downarrow$ is the spin density and
$\rho_{st}=\rho_{n\uparrow}-\rho_{n\downarrow}-\rho_{p\uparrow}+\rho_{p\downarrow}$ is the spin-isospin density.  The
spin symmetry is satisfied if the power of the density dependent
terms $\gamma_s$ and $\gamma_{st}$ are both  even integers.

In the following study,  the spin-density dependent
terms~(\ref{eq:addint}) are added to  the original Hamiltonian. Then
the density dependent part of the Skyrme energy density functional,
\bea
 \mathcal{H}_3=  \frac{t_3}{48}
\rho^{\alpha}\big[3\rho^2+(2x_3-1)\rho_s^2-(2x_3+1)\rho_t^2-\rho_{st}^2
\big] , \eea has the extra density dependent  terms
$\mathcal{H}_3^{s}$ and $\mathcal{H}_3^{st}$ which read, \bea
\mathcal{H}_3^s=\frac{t_3^{s}}{48}
\rho_{s}^{\gamma_{s}}\big[3\rho^2+(2x_3^{s}-1)\rho_s^2-(2x_3^{s}+1)\rho_t^2-\rho_{st}^2
\big] , \\
\mathcal{H}_3^{st}=\frac{t_3^{st}}{48}
\rho_{st}^{\gamma_{st}}\big[3\rho^2+(2x_3^{st}-1)\rho_s^2-(2x_3^{st}+1)\rho_t^2-\rho_{st}^2
\big] , \eea where 
$\rho_{t}=\rho_{n}-\rho_{p}$. The  mean field potential $U_q$, where
$q=n,p$, gets additional  terms \bea
U_q^\mathrm{add.}&=&\frac{t_3^s}{12}\rho_s^{\gamma_s}\big
[(2+x_3^s)\rho-(1+2x_3^s)\rho_q\big ]
\nonumber \\
&+&\frac{t_3^{st}}{12}\rho_{st}^{\gamma_{st}}\big
[(2+x_3^{st})\rho-(1+2x_3^{st})\rho_q\big ]. \label{eq:uq} \eea

In symmetric nuclear matter the Landau parameters~\cite{Nav97} are
also modified by the following additional terms
\bea
\frac{F_0^\mathrm{add.}}{N_0}
&=&\frac{t_3^{s}}{8}\rho_s^{\gamma_{s}}+\frac{t_3^{st}}{8}\rho_{st}^{\gamma_{st}}
 \; , \label{eq:f0sm}\\
\frac{F_0^{\prime \mathrm{add.}}}{N_0}
&=&-\frac{t_3^{s}}{24}(2x_3^{s}+1)\rho_s^{\gamma_{s}}
-\frac{t_3^{st}}{24}(2x_3^{st}+1)\rho_{st}^{\gamma_{st}} \; , \label{eq:fp0sm}\\
\frac{G_0^\mathrm{add.}}{N_0}  &=& \frac{t_3^s}{48}\gamma_s(\gamma_s-1)
[3\rho^2-(2x_3^s+1)\rho_t^2\nonumber \\
&-&\rho_{st}^2 ]\rho_s^{\gamma_s-2}
+\frac{t_3^{st}}{12}(x_3^{st}-\frac{1}{2})\rho_{st}^{\gamma_{st}}
\nonumber \\
&+&\frac{t_3^s}{24}(x_3^s-\frac{1}{2})\left(\gamma_s+1\right)\left(\gamma_s+2\right)\rho_s^{\gamma_s}
 \; , \label{eq:g0sm}\\
\frac{G_0^{\prime \mathrm{add.}}}{N_0} &=&
\frac{t_3^{st}}{48}\gamma_{st}(\gamma_{st}-1)
[3\rho^2+\left(2x_3^{st}-1\right)\rho_s^2
\nonumber \\
&-&(2x_3^{st}+1)\rho_t^2 ]\rho_{st}^{\gamma_{st}-2}
-\frac{t_3^s}{24}\rho_s^{\gamma_s}
\nonumber \\
&-&\frac{t_3^{st}}{48}(\gamma_{st}+2)(\gamma_{st}+1)\rho_{st}^{\gamma_{st}} \; .\label{eq:gp0sm}
\eea

It was mentioned in ~\cite{marg1,marg2}  that the spin-density
dependent terms may lead  very important effects on  the spin and
the spin-isospin properties of finite nuclei and nuclear matter.
That is,  the dimensionless Landau parameter $G_0^\prime$ is
increased by about 0.3 for three interactions SLy5, LNS and BSk16.
The improved Skyrme energy density functional in spin and
spin-isospin channels will  give also substantial contributions to
the spin and spin-isospin excitations in finite nuclei, such as M1
and GT excitations. In present work, we will study the effect of
spin-density dependent terms on the spin-dependent M1 and GT
excitations in finite nuclei $^{90}$Zr and $^{208}$Pb.

The calculations are done within the Skyrme HF plus RPA. The well
known RPA method~\cite{Ring80,Rowe} in matrix form is given by
\bea\label{RPA} \left( \begin{array}{cc}
  A & B \\
B^* & A^* \end{array}  \right) \left( \begin{array}{c}
X^\nu \\
Y^\nu  \end{array} \right) =E_\nu \left( \begin{array}{cc}
1 & 0\\
0 & -1 \end{array}  \right) \left( \begin{array}{c}
X^\nu \\
Y^\nu  \end{array} \right), \eea where $E_\nu$ is the energy of the
$\nu$-th RPA state and  X$^\nu$, Y$^\nu$ are the corresponding
forward and backward amplitudes, respectively. The matrix elements
$A$ and $B$ are expressed as
\begin{eqnarray}
A_{mi,nj}=(\epsilon_m-\epsilon_n)\delta_{mn}\delta_{ij}+\langle{mj|V_{res}|in}\rangle,
\end{eqnarray}
\begin{eqnarray}
B_{mi,nj}=\langle{mn|V_{res}|ij}\rangle.
\end{eqnarray}

The p-h matrix elements are obtained from the Skyrme energy density
functional including all the terms in Eqs. (3)$\sim$(5). The
explicit forms of the matrices $A$ and $B$  are given in
Ref.~\cite{colo13} in the case of Skyrme force. In general, the
expression of the residual interaction is derived from the second
derivative of the energy density with respect to the density
$\rho_{st}$ with the spin abd ispspin  indices,
\begin{equation}
V_{res}=\sum_{sts^\prime t^\prime}
\frac{\delta^2H}{\delta\rho_{st}\delta\rho_{s^\prime t^\prime}}
,\label{res}
\end{equation}
where $H$ is the HF energy density functional. According to
Eq.~\eqref{res}, the antisymmetrized particle-hole interaction
induced by the spin-density dependent terms \eqref{eq:addint} are
expressed as,
\begin{eqnarray}
V_{res}^{\rm qq} & = & v_{0}^{\rm qq}\delta\left( \vec r_1 - \vec
r_2 \right) + v_{\sigma}^{\rm qq}\delta\left( \vec r_1 - \vec r_2
\right)
{\bm\sigma}_1\cdot{\bm\sigma}_2, \nonumber \\
V_{res}^{\rm qq'} & = & v_{0}^{\rm qq'}\delta\left( \vec r_1 - \vec
r_2 \right) + v_{\sigma}^{\rm qq'}\delta\left( \vec r_1 - \vec r_2
\right) {\bm\sigma}_1\cdot{\bm\sigma}_2,
\end{eqnarray}
where the functions $v_0$ and $v_\sigma$ depend only on the radial coordinate $r$ and their
detailed expressions are given by
\begin{eqnarray}
v_{0}^{\rm qq}(r) &
=&-\frac{t_3^{s}}{12}(x_3^{s}-1)\rho_s^{\gamma_{s}}
-\frac{t_3^{st}}{12}(x_3^{st}-1)\rho_{st}^{\gamma_{st}}
\nonumber\\
v_{0}^{\rm qq'}(r) & =
&\frac{t_3^{s}}{12}(x_3^{s}+2)\rho_s^{\gamma_{s}}
+\frac{t_3^{st}}{12}(x_3^{st}+2)\rho_{st}^{\gamma_{st}}
\nonumber\\
v_{\sigma}^{\rm qq}(r) & =
&\frac{t_3^{s}}{48}\bigg[\gamma_s(\gamma_s-1)\rho_s^{\gamma_s-2}
\bigg(3\rho^2-(2x_3^{s}+1)\rho_t^2-\rho_{st}^2\bigg) \nonumber\\
&&+\rho_s^{\gamma_s}\bigg((\gamma_s+1)(\gamma_s+2)(2x_3^{s}-1)-2\bigg)\bigg]
\nonumber\\
&&+\frac{t_3^{st}}{48}\bigg[\gamma_{st}(\gamma_{st}-1)\rho_{st}^{\gamma_{st}-2}
\bigg(3\rho^2+(2x_3^{st}-1)\rho_s^2\nonumber\\&&-(2x_3^{st}+1)\rho_t^2\bigg)
\nonumber\\
&&+\rho_{st}^{\gamma_{st}}\bigg(-(\gamma_{st}+1)(\gamma_{st}+2)+
2(2x_3^{st}-1)\bigg)\bigg],
\nonumber \\
v_{\sigma}^{\rm qq'}(r) & =
&\frac{t_3^{s}}{48}\bigg[\gamma_s(\gamma_s-1)\rho_s^{\gamma_s-2}
\bigg(3\rho^2-(2x_3^{s}+1)\rho_t^2-\rho_{st}^2\bigg)\nonumber\\&&
+\rho_s^{\gamma_s}\bigg((\gamma_s+1)(\gamma_s+2)(2x_3^{s}-1)+2\bigg)\bigg]
\nonumber\\
&&+\frac{t_3^{st}}{48}\bigg[-\gamma_{st}(\gamma_{st}-1)\rho_{st}^{\gamma_{st}-2}
\bigg(3\rho^2+(2x_3^{st}-1)\rho_s^2\nonumber\\&&-(2x_3^{st}+1)\rho_t^2\bigg)
\nonumber\\
&&+\rho_{st}^{\gamma_{st}}\bigg((\gamma_{st}+1)(\gamma_{st}+2)+
2(x_3^{st}-1)\bigg)\bigg]
\end{eqnarray}

We will use the
following  operator for  M1 excitation,
\begin{eqnarray}
\hat{F}_{M1}=\sum^A_{i=1}\{g^s_i\overrightarrow{s_i}+g^l_i\overrightarrow{l}\},
\end{eqnarray} where the spin $g-$factors are  $g^s=5.586$ for protons and $g^s=-3.826$ for
neutrons, respectively, and the orbital  $g-$factors are  $g^l=1.0$
for protons and $g^l$=0.0 for neutrons, respectively,  in unit of
the nuclear magneton $\mu_N=e\hbar/2mc$.  We will also study  the
charge-exchange GT  excitations. The  GT external operator reads
\begin{eqnarray}
\hat{F}_{GT\pm}=\sum^A_{i=1}\overrightarrow{\sigma}(i)t_{\pm}(i).
\end{eqnarray}

\section{Results and discussions}

In Table~I we show the parameters used in this study and the Landau
parameters $G_0$ and $G_0^\prime$ calculated with the corresponding
Skyrme interactions. To keep the spin symmetry, we set $\gamma_s$
and $\gamma_{st}$ equal to 2. The values for the other parameters
$t_3^s$, $t_3^{st}$, $x_3^s$ and $x_3^{st}$ are fixed by an optimal
fit of the BHF results in spin and spin-isospin channels in a higher
density region than the normal density. The ground state properties
of nuclei $^{90}$Zr and $^{208}$Pb are calculated in the coordinate
space with a box approximation. The radius of the box is taken to be
20\,fm in which  the continuum is discretized  in  the large  box.
The 8 oscillator shell is included as  the particle states to build
the RPA model space. All calculations are performed within the SLy5
parameter set by including or excluding the spin-density dependent
terms.
\begin{table}[tp]
\setlength{\tabcolsep}{.05in}
\renewcommand{\arraystretch}{1.3}
\caption{\label{tab1} Parameters of the spin-density dependent terms
$t_3^s$ (in MeV.fm$^{3\gamma_s-2}$), $t_3^{st}$ (in
MeV.fm$^{3\gamma_{st}-2}$), $x_3^s$ and  $x_3^{st}$ for the
interactions SLy5st. We also show the dimensionless  Landau
parameters $G_0$ and $G_0^\prime$ deduced from the original  SLy5
and  SLy5st, respectively.}

\begin{tabular}{ccccccccc}
\hline
& $t_3^s$ & $t_3^{st}$ & $x_3^s$ & $x_3^{st}$ & $\gamma_{s}$ & $\gamma_{st}$ & $G_0$ &  $G_0^\prime$ \\
\hline
SLy5 & - & - & - & - & - & - & 1.12 & $-0.14$ \\
SLy5st & 0.6$\times$10$^4$ & 2$\times$10$^4$ &  -3  & 0 & 2 & 2  & 1.19  & 0.15\\
\hline
\end{tabular}
\end{table}

\begin{figure}[tb]
\begin{center}
\includegraphics[width=0.9\linewidth]{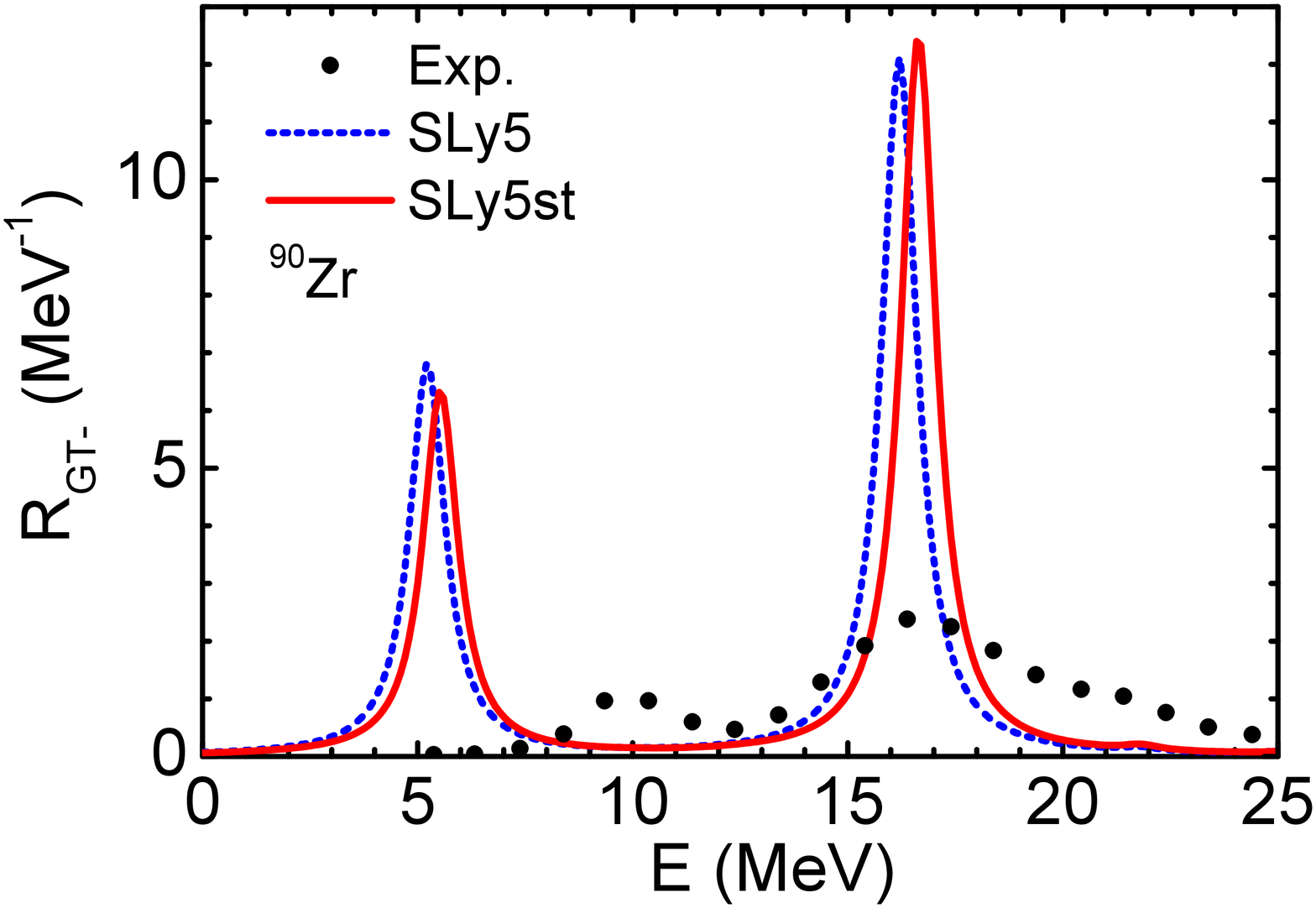}
\vglue -2.5cm
\includegraphics[width=0.9\linewidth]{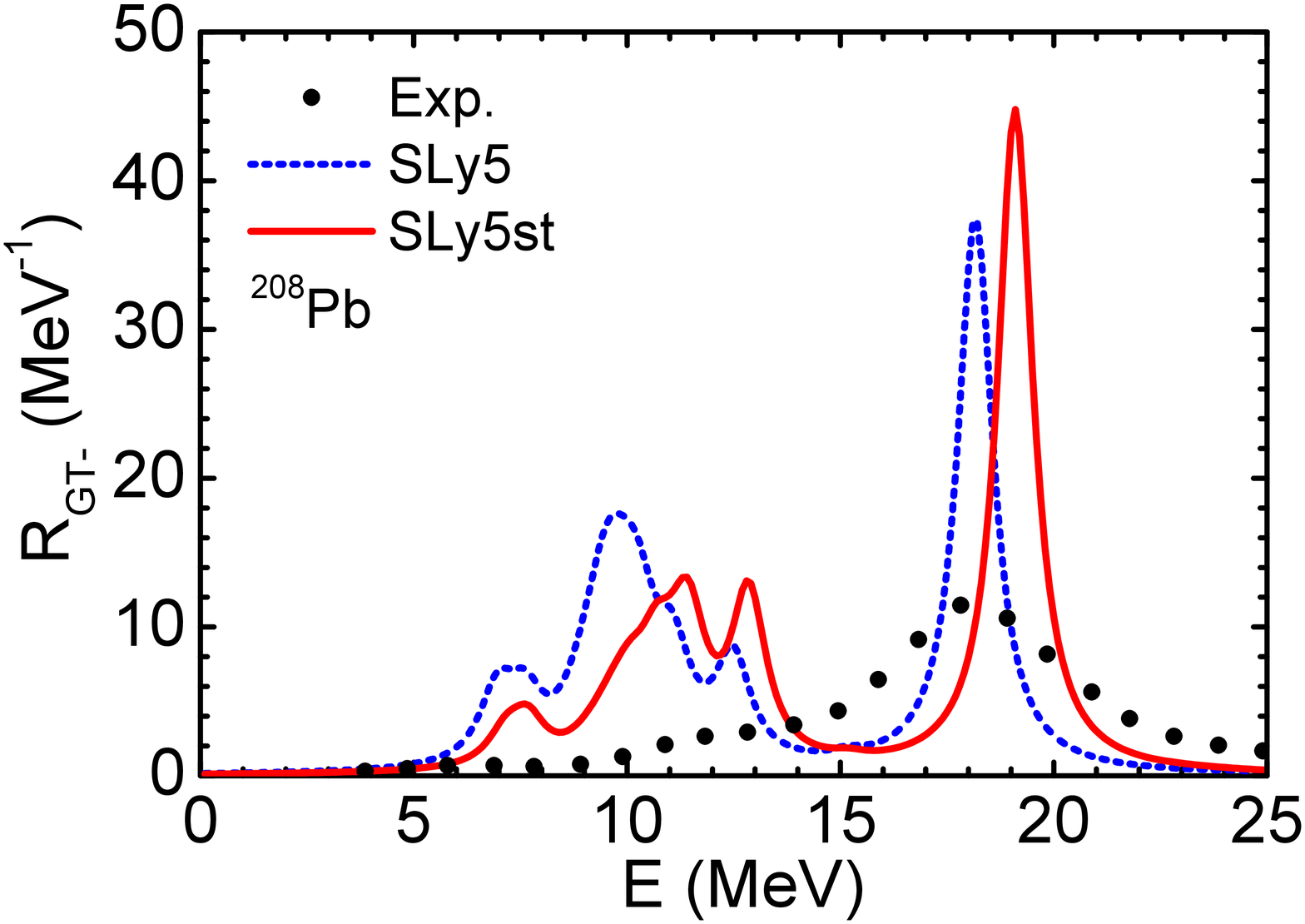}
\end{center} \vglue -3.cm\caption{(Color online) RPA Response functions of
$^{90}$Zr (upper panel) and $^{208}$Pb (lower panel) for GT
excitations calculated by the Skyrme HF plus RPA approach based on
the SLy5 interaction. Solid (Dotted)  line is the result given by
including (excluding) the spin-density dependent terms. A
Lorentzian smearing parameter equals 1 MeV. The experimental
responses from Ref.~\cite{Wak12,Wak97} are shown by the dots.}
\label{fig1}
\end{figure}

In Fig.1 we display the response functions for GT excitation in
$^{90}$Zr and $^{208}$Pb calculated with and without
  the contribution of the spin-density dependent terms~\eqref{eq:addint}.
The solid (dotted) line represents the results including (excluding)
the spin-density dependent terms  both at the HF  and RPA
calculations. The dots show  the corresponding experimental GT
response. As one can see from Fig.1, the inclusion of the
spin-density dependent terms  tends to slightly increase the
high-lying strength on the one hand and to decrease the low-lying
strength on the other hand. The excitation  energies are shifted up
in energy both the low-lying and high-lying strengths by the
spin-dependent terms. Without the spin-density dependent terms, the
centroid energies of the low-lying and high-lying strengths are 5.23
MeV (9.87 MeV) and 16.26 MeV (18.13 MeV) for $^{90}$Zr ($^{208}$Pb).
Including the spin-density dependent terms, the centroid energies of
the low-lying and high-lying strengths become  5.53 MeV (10.92 MeV)
and 16.68 MeV (19.07 MeV) for $^{90}$Zr ($^{208}$Pb).
 The energy shift is  0.3 MeV
(1.05 MeV) for the low-lying and 0.42 MeV (0.94 MeV) for the
high-lying states in $^{90}$Zr ($^{208}$Pb).
The energy shift given by the Skyrme HF plus RPA calculations 
 is qualitatively the same as
 those  estimated by the semi-classical Steinwedel-Jensen
model for $^{208}$Pb  in Ref.~\cite{marg1}. The upward shift of the
centroid energies can be understood as follows:  the spin-density
dependent terms give an strong repulsive contribution to the matrix
elements of  RPA  for  the GT calculations because the residual
interactions or the Landau parameter $G'_0$ changes  to be more positive
from $-0.14$ to 0.15  when the spin-density dependent terms are
included. It also can been seen that the RPA collective state
located at 19.07 MeV with the spin-density dependent terms in
$^{208}$Pb is very close to the experimental GT excitation energy of
19.2 $\pm$ 0.2 MeV~\cite{Aki95,Wak12}. For $^{90}$Zr, the calculated
values with or without the contribution of the spin-density
dependent terms both are larger than the experimental value of 15.60
MeV~\cite{Wak97}.

\begin{figure}[htb]
\begin{center}
\includegraphics[width=0.9\linewidth]{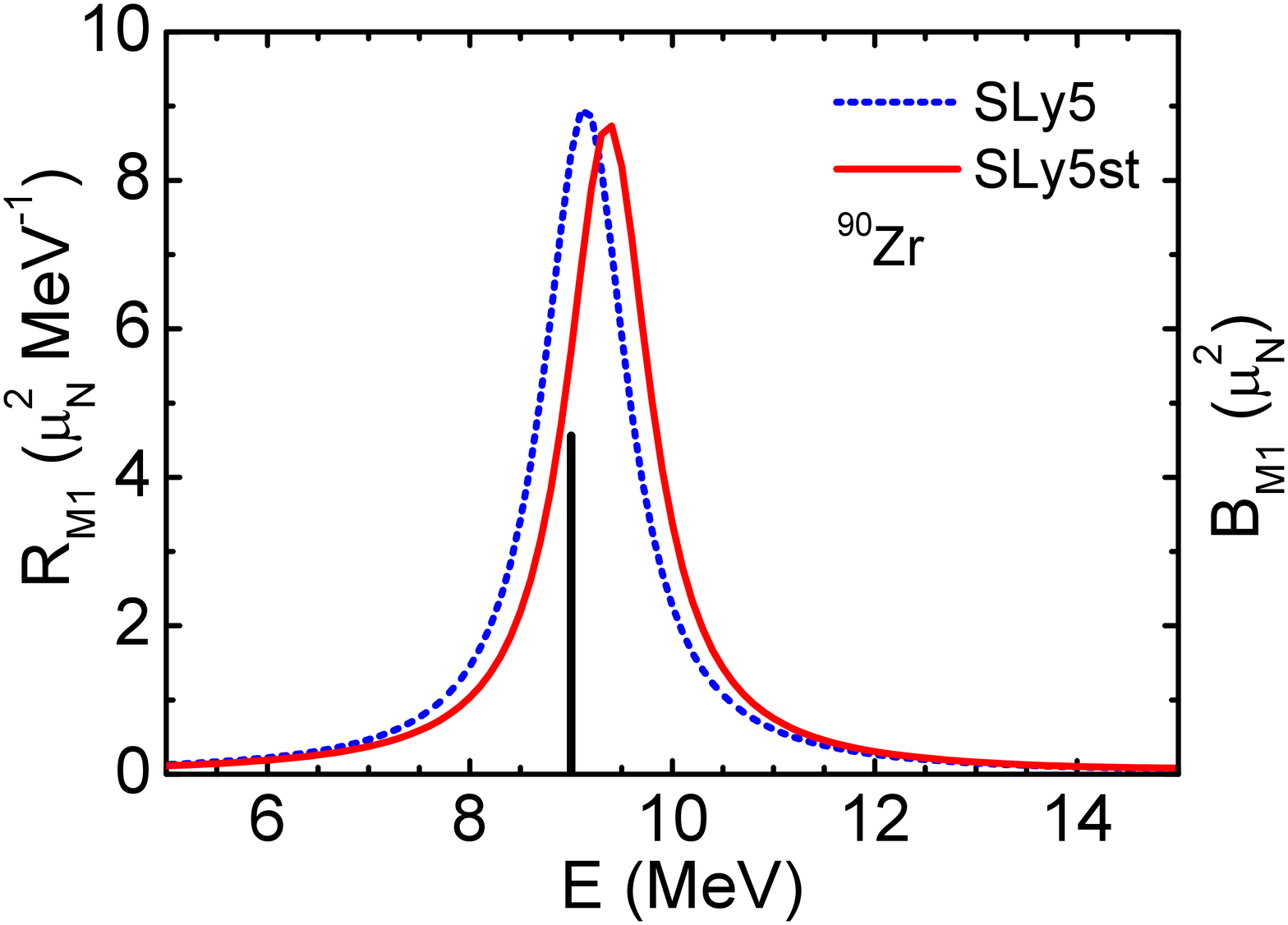}
\vglue -2.5cm
\includegraphics[width=0.9\linewidth]{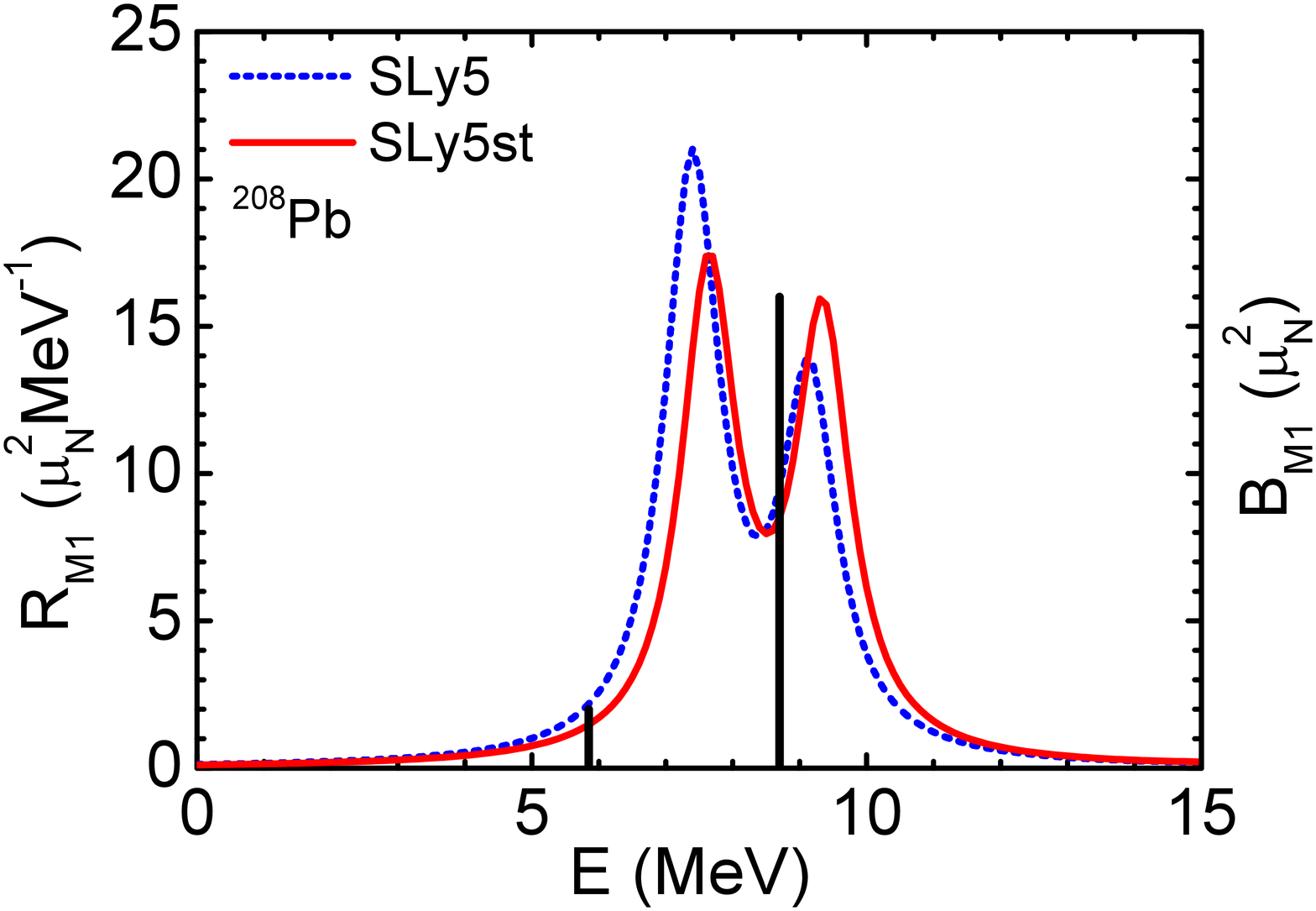}
\end{center} \vglue -3.cm\caption{(Color online) RPA Response functions of
$^{90}$Zr (upper panel) and $^{208}$Pb (lower panel) for M1
excitations calculated by the Skyrme HF plus RPA approach based on
the SLy5 interaction. Solid (Dotted)  line is the result given by
including ( excluding) the spin-density dependent terms. A
Lorentzian smearing parameter equals 1 MeV. The experimental B(M1)
values from Ref.~\cite{Shi08,Las88,Koh87,Rus13,Iwa12,Las87} are
shown by the bars.} \label{fig1}
\end{figure}

We also investigate the effect of the extended Skyrme interaction on
the M1  excitation by  the RPA calculations for $^{90}$Zr and
$^{208}$Pb. The results are shown in Fig. 2. The spin-density
dependent terms are included or excluded in the calculations to
clarify  their influence. For the RPA results of $^{208}$Pb, the
proton 1h$_{11/2}\rightarrow$ 1h$_{9/2}$ configuration contributes
mainly to the lower energy peak of the response function, and the
neutron 1i$_{13/2}\rightarrow$ 1i$_{11/2}$ configuration plays the
main role in the higher energy peak. For $^{90}$Zr, the M1 response
function mainly comes from the neutron configuration of
1g$_{9/2}\rightarrow$ 1g$_{7/2}$. The results show that the
inclusion of the spin-density dependent terms increases the energies
of M1 states  both in $^{90}$Zr and $^{208}$Pb. The peak energies of
M1 excitation in $^{208}$Pb are 7.6 MeV (7.4 MeV) for the lower one
and 9.3 MeV (9.1 MeV) for the higher one with including (excluding)
the spin-density dependent terms. The peak energy of the M1
excitation in $^{90}$Zr are 9.4 MeV (9.1 MeV) by including
(excluding) the spin-density dependent terms. The energy shift is
less than 0.3 MeV for both $^{90}$Zr and $^{208}$Pb. The effect  of
the spin-dependent terms is predicted to be smaller on the
distribution of the M1 response function compared with that on the
GT excitation. This can be understood by the change of the Landau
parameters when the spin-density dependent terms are included. The
difference of $G_0$ which contributes to the M1 excitation is about
0.07 (from 1.12 to 1.19),  while the change of $G_0^\prime$ which
plays the dominant role in GT excitation is about 0.3 (from -0.14 to
0.15). In Fig. 2 we  show also the experimental data of the M1
excitations  for $^{90}$Zr and $^{208}$Pb. The experimental data for
the M1 excitations in $^{208}$Pb are found at Ex=5.85 MeV for
low-lying component and  between Ex=7.1 and 8.7 MeV for the
high-lying component~\cite{Shi08,Las88,Koh87}, while in $^{90}$Zr
the M1 strengths exist between Ex= 9.0 and 9.53
MeV~\cite{Rus13,Iwa12,Las87}. We can see that the present
theoretical results, taking or not taking into account the
contribution of the spin-density dependent terms, slightly
overestimate the experimental data in energy.

\section{SUMMARY AND PERSPECTIVE}

In summary, we have studied the effect of the spin-density dependent
terms of the Skyrme energy density functional on the M1 and GT giant
excitations in $^{90}$Zr and $^{208}$Pb by  the Skyrme HF plus RPA
calculations. The calculations are carried out with the  SLy5 Skyrme
interaction and the extended Skyrme interaction  SLy5st, in which
the spin-density dependent terms are added to the SLy5 parameter set
to mimic the BHF results in spin and spin-isospin channels. Those
terms are switched on and off in both the HF and the RPA
calculations in this study. The inclusion of spin-density dependent
terms is known to give no contribution to the ground state of
even-even nuclei, while the residual interactions  from the
spin-density dependent terms give  substantial repulsive effect and
shifts the M1 and GT response function of finite nuclei to higher
energy.

The main conclusion we can draw from the present study is that the
spin and spin-isospin response functions can be changed without
altering the ground-state properties. Since the parameters related
to the spin-density dependent terms are introduced to the existing
Skyrme interaction, it is marginal whether the new interaction
improves the agreement with the experimental data of spin dependent
excitations or not. The better strategy  could be  to perform a
global fitting of the parameters of the spin-independent and
spin-dependent density terms on the same foot. Recently,
 the effect of tensor force on
the various response of nuclear systems has been studied
extensively~\cite{Cao11,Bai11,Pas121} and the  important
contributions to the spin and spin-isospin response in nuclear
matter and finite nuclei are pointed out. It is a future challenge
 to include both the spin dependent terms and the
tensor force in the parameter fit procedure.   This study will be discussed
in  the forthcoming paper.

\section*{ACKNOWLEDGEMENTS}
This work is supported by the National Natural Science Foundation of
China under Grant Nos 10875150 and 11175216, and is supported in
part by the Project of Knowledge Innovation Program (PKIP) of
Chinese Academy of Sciences, Grant No. KJCX2-EW-N01. This work is
partially supported by the Japanese Ministry of Education, Culture,
Sports, Science and Technology by Grant-in-Aid for Scientific
Research under the program number (C(2))20540277.

\end{CJK*}

\end{document}